\documentclass{llncs}
\usepackage{amssymb}
\usepackage{amsmath}
\usepackage{graphicx}
\newcommand{\set}[1]{\left\{#1\right\}} % Set of elements
\newcommand{\staten}[1]{\Omega_{#1}} % Number of microstates at E
\def\Z{\mathbb{Z}}
\def\lat{\Gamma}
\def\autlat{\mathrm{Aut}\left(\lat\right)}
\def\latv{V(\lat)}
\def\latvn{\left|V(\lat)\right|} % Number of lattice verteces
\def\stateset{\Sigma} % Set of (multi,micro)states
\def\tra{\mathbf{T}^2}
\def\dih{\mathbf{D}_4}
\begin{document}
%\ifpdf
\title{Symmetries and Dynamics of Discrete Systems}
\titlerunning{Symmetries and Dynamics of Discrete Systems}
\author{Vladimir V. Kornyak}
\institute{Laboratory of Information Technologies \\
           Joint Institute for Nuclear Research \\
           141980 Dubna, Russia \\
           \email{kornyak@jinr.ru}}
\authorrunning{Vladimir V. Kornyak}
\maketitle
\begin{abstract}
We consider discrete dynamical systems and lattice models 
in statistical mechanics from the point of view of their
symmetry groups.
We describe a C program for symmetry analysis of discrete systems.
Among other features, the program constructs and investigates 
\emph{phase portraits} of discrete dynamical systems \emph{modulo 
groups} of their symmetries, searches dynamical systems 
possessing specific properties, e.g.,\emph{reversibility},
computes microcanonical \emph{partition functions} and searches 
\emph{phase transitions} in mesoscopic systems. 
Some computational results and observations are presented.
In particular, we explain formation of moving soliton-like structures 
similar to ``\emph{spaceships}'' in cellular automata.
\end{abstract}
\section{Introduction}
Symmetry analysis of continuous systems described by ordinary or partial differential
equations is well developed and fruitful discipline. But there is a sense of 
incompleteness of the approach since the transformations
used in the symmetry analysis of continuous systems --- point and contact Lie, 
B\"acklund and Lie--B\"acklund,  
some sporadic instances of so-called
\emph{non-local} transformations ---  constitute negligible small part of all thinkable transformations.
In this context finite discrete systems look more attractive since we can study 
\emph{all possible} their symmetries.
\par
Furthermore, there are many hints from quantum mechanics and quantum gravity that 
discreteness is more suitable for describing physics at small distances than continuity 
which arises only as a logical limit in considering large collections of discrete structures. 
\par
Both differential equations and cellular atomata are based on the idea of \emph{locality} --- behavior of a system as a whole is determined by interections of its closely situated parts.
Recently \cite{KornyakCASC05,KornyakProg06} we  
showed that any collection
of discrete points taking values in finite sets possesses some kind of locality.
More specifically, let us consider collection  of 
$N$ ``points'',  symbolically $\delta = \set{x_1,\ldots,x_N}$.
We call $\delta$ \emph{domain}. Each $x_i$ takes value in its own set of values
$Q_i = \set{s^1_i,\ldots,s^{q_i}_i}$ or using the standard notation
$Q_i = \set{0,\ldots,q_i-1}$. Adopting $Q^\delta$ as symbolical notation for 
the Cartesian product $Q_1\times\cdots\times Q_N$, we define \emph{relation} 
on $\delta$ as an arbitrary subset $R^\delta \subseteq Q^\delta$. 
Then we define \emph{consequence} of relation $R^\delta$ as an \emph{arbitrary} 
superset $S^\delta\supseteq R^\delta$ and \emph{proper consequence}
as a consequence which can be represented in the form $P^\alpha\times Q^{\delta\setminus\alpha}$,
where $P^\alpha$ is \emph{nontrivial} (i.e., $P^\alpha \neq Q^\alpha$) relation 
on the proper subset $\alpha\subset\delta$. We show that any relation $R^\delta$
allows a decomposition in terms of its proper consequences. This decomposition naturally
imposes a structure of \emph{abstract simplicial complex} --- one of 
the mathematical abstractions of locality. Thus we call collections of discrete finite-valued
points \emph{discrete relations on abstract simplicial complexes}.
\par
We demonstrated also that such relations in special cases
correspond to \emph{systems of polynomial equations} 
(if all points $x_i$ take values in the same set $Q$ and its cardinality 
is a power of a prime $\left|Q\right|=p^k$) and to \emph{cellular automata} 
(if domain $\delta$ allows decomposition into congruent simplices with 
the same relation on the simplices and this
relation is \emph{functional}). 
The notion of discrete relations covers also discrete dynamical systems more general than 
cellular automata. 
The lattice models in statistical mechanics can also be included in this framework by considering
\emph{ensembles} of discrete relations on abstract simplicial complexes.
\par
In this paper we  study dependence of behavior of discrete dynamical systems on graphs 
--- one-dimensional simplicial complexes --- on symmetries of the graphs.
We describe our C program for discrete symmetry analysis and results of 
its application to cellular automata and mesoscopic lattice models.
\section{Symmetries of Lattices and Functions on Lattices}
\subsubsection{Lattices.}
A space of discrete dynamical system will be called a \emph{lattice}.
Traditionally, the word `lattice' is often applied to some regular system of separated points 
of a continuous metric space. In many problems of applied mathematics and mathematical physics
both metrical relations between discrete points and existence of underlying continuous manifold
do not matter. The notion of `adjacency' for pairs of points is essential only. 
All problems considered in the paper are of this kind. Thus we define a \emph{lattice} 
as indirected $k$-regular graph $\lat$ without loops and multiple edges whose \emph{automorphism group}
 $\autlat$ acts transitively on the set of vertices  $\latv$.
Sometimes we shall depict our lattices as embedded in some continuous spaces like spheres or tori
(in this case we can talk about `dimension' of lattice). But such representations 
are not significant in our context and used only for vizualization.
\par
\begin{figure}[!h]
%\vspace*{-10pt}
\centering
\includegraphics[width=0.9\textwidth]{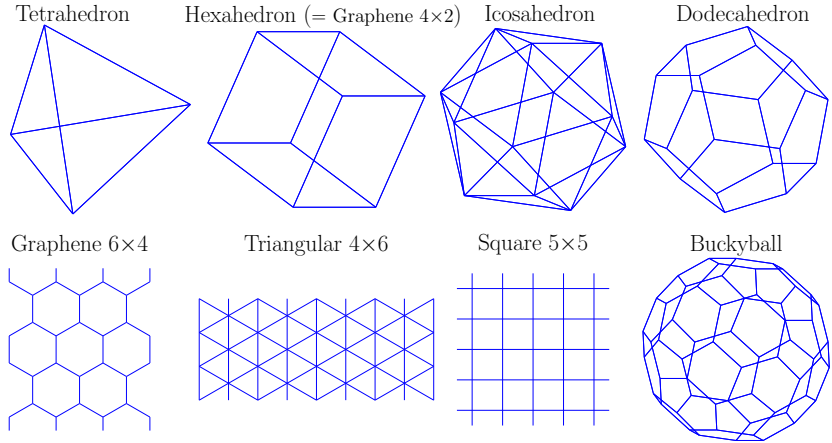}
\caption{Examples of lattices}
	\label{lattices}
\end{figure}

 
The lattices we are concerned in this paper are shown in Fig. \ref{lattices}.
%\par
Note that the lattices marked in Fig. \ref{lattices} as ``Graphene 6$\times$4'', 
``Triangular 4$\times$6'' and ``Square 5$\times$5'' can be closed by identifications 
of opposite sides of rectangles in several different ways. Most natural identifications form regular graphs 
embeddable in the torus and in the Klein bottle.
Computation shows that the Klein bottle arrangement (as well as others except for embeddable in 
the torus) leads to \emph{nonhomogeneous} lattices.
For example, the hexagonal lattice ``Graphene 6$\times$4'' embeddable in the Klein bottle has 
16-element symmetry group and this group splits the set of vertices into two orbits of sizes 8 and 16.
Since non-transitivity of points contradicts to our usual notion of space (and our definition of lattice), 
we shall not consider further such lattices.
\par
\begin{figure}%[!h]
\centering
\includegraphics[width=0.8\textwidth]{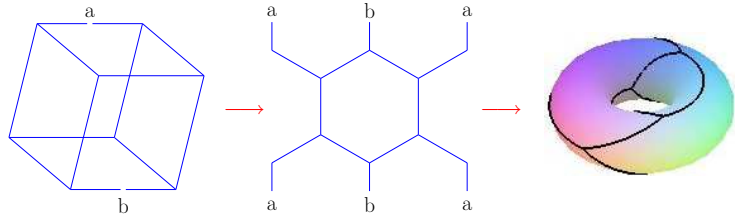} 
\caption{The same graph forms 4-gonal (6 tetragons) lattice in sphere
$\bbbs^2$ and 6-gonal (4 hexagons) lattice in torus $\bbbt^2$.}
	\label{Cube-on-tor}
\end{figure}

It is interesting to note that the graph of hexahedron can be interpreted -- as is clear
from Fig. \ref{Cube-on-tor}  -- either as 4-gonal lattice in sphere or as 6-gonal lattice in torus.
\par
\subsubsection{Computing Automorphisms.}
The automorphism group of graph with $n$ vertices may have up to $n!$ elements.
However, McKay's algorithm \cite{McKay}, based on efficiently arranged search tree, 
determines the graph automorphisms by constructing small number of the group generators.
This number is bounded by $n-1$, but usually it is much less.
\par
In Sect. \ref{solitonformation} we discuss the connection of formation of soliton-like 
structures in discrete systems with symmetries of lattices. 
There we consider concrete example of system on square lattice.
So let us describe symmetries of $N\times N$ square lattices in more detail. 
We assume that the lattice has valency 4 (``von Neumann neighborhood'') or 8 (``Moore neighborhood''). 
We assume also that the lattice is closed into discrete torus $\Z_N\times\Z_N$, if $N < \infty$.
Otherwise the lattice is discrete plane  $\Z\times\Z$. 
In both von Neumann and Moore cases the symmetry group, which we denote by $G_{N\times N}$, 
is the same.
The group has the structure of \emph{semidirect} product of the subgroup 
of \emph{translations} $\tra = \Z_N\times\Z_N$ 
(we assume $\Z_\infty = \Z$) and \emph{dihedral group} $\dih$
\begin{equation}
	G_{N\times N} = \tra\rtimes\dih,\mbox{~~if~~} N = 3,5,6,\ldots,\infty.
	\label{normalgnn}
\end{equation}
The dihedral group $\dih$
is, in its turn, the semidirect product  $\dih = \Z_4\rtimes\Z_2$. 
Here $\Z_4$ is generated by $90^o$  rotations, and $\Z_2$ are reflections.
The size of $G_{N\times N}$ is 
$$\left|G_{N\times N}\right| = 8N^2,\mbox{~~if~~} N\neq4.$$
%\par
In the case $N=4$ the size of the group becomes three times larger than expected 
$$\left|G_{4\times4}\right| = 3\times8\times4^2 \equiv384.$$
This anomaly results from additional $\Z_3$ symmetry in the group  $G_{4\times4}$. 
Now the translation subgroup $\tra = \Z_4\times\Z_4$ is \emph{not normal} 
and the structure of $G_{4\times4}$ differs essentially from (\ref{normalgnn}). 
The algorithm implemented in the computer algebra system \textbf{GAP} \cite{GAP} 
gives the following structure
\begin{equation}
G_{4\times4} = \overbrace{	\left(\left(\left(\left(\Z_2 \times \dih\right) 
\rtimes \Z_2\right) \rtimes \Z_3\right) \rtimes \Z_2\right)}^{\mbox{normal
closure of~~}{\textstyle \tra}} \rtimes \Z_2 .
\end{equation}
\subsubsection{Functions on Lattices.}
To study the symmetry properties of a system 
on a lattice $\lat$ we should consider action of the group $\autlat$ on
the space $\stateset = Q^\lat$ of $Q$-valued functions on 
$\lat$, where $Q =\left\{0,\ldots,q-1\right\}$
is the set of values of lattice vertices. 
We shall call the elements of $\stateset$ \emph{states} or 
(later in Sect. \ref{mesosect}) \emph{microstates}.
\par
The group  $\autlat$ acts
non-transitively on the space $\stateset$ splitting  
this space into the disjoint orbits of different sizes
$$
\stateset = \bigcup\limits_{i=1}^{N_{orbits}}O_i\enspace.
$$
The action of $\autlat$ on $\stateset$ is defined by
$$\left(g\varphi\right)\left(x\right) = \varphi\left(g^{-1}x\right)\ ,$$
where $x\in\latv,~\varphi\left(x\right)\in\stateset,~g\in\autlat$.\\
Burnside's lemma counts the total number of orbits in the state space $\stateset$ 
$$
N_{orbits} = \frac{1}{\left|\autlat\right|}
\sum\limits_{g\in\autlat}q^{N_{cycles}^g}\enspace.
$$
Here $N_{cycles}^g$ is the number of cycles in the group element $g$.
\par
\noindent
Large symmetry group 	allows to represent dynamics on the lattice in more compact form. 
For example, the automorphism group of (graph of) icosahedron, dodecahedron and buckyball
is $\mathrm{S}_5$%
\footnote{Traditionally, the icosahedral group $\mathrm{I}_h = \mathrm{A}_5$ is adopted 
as a symmetry group for these polyhedra. $\mathrm{A}_5$ is 60-element discrete subgroup of $\mathrm{SO}(3)$.
Adding reflections to $\mathrm{A}_5$ we get twice larger (and hence more efficient for our 
purposes) group $\mathrm{S}_5$.},
and the information about behavior 
of any dynamical system on these lattices can be compressed nearly in proportion to $\left|\mathrm{S}_5\right|=120$.
\subsubsection{Illustrative Data.}
In Table \ref{LatAutOrbs} we collect some quantitative information about the lattices from
Fig.\ref{lattices} and their automorphism groups, namely, \emph{number of vertices}  $\latvn$,
\emph{size of automorphism group} $\left|\autlat\right|$, total \emph{number of states} 
$\staten{} = \left|\stateset\right|\equiv q^{\latvn}$ (assuming $q=2$) and \emph{number of group orbits} $N_{orbits}$ in the space of states.  
\begin{table}[htbp]
	\centering
	\caption{Lattices, groups, orbits:
	 quantitative characteristics.}
	\label{LatAutOrbs}
		\begin{tabular}{l|c|c|c|c}
Lattice&$\latvn$&$\left|\autlat\right|$&$\staten{}=q^{\latvn}$&$N_{orbits}$
\\\hline
Tetrahedron& 4  & 24 & 16  & 5
\\\hline
Hexahedron  &	 8  & 48   &  256 & 22
\\\hline
Icosahedron &	12 & 120  & 4096  & 82
\\\hline
Dodecahedron&	20 & 120  & 1048576  & 9436
\\\hline
\begin{tabular}{l}
\!Graphene 6$\times$4\\
\!Torus
\end{tabular} 
&	24 & 48 & 16777216& 355353
\\\hline
\begin{tabular}{l}
\!Graphene 6$\times$4\\
\!Klein bottle
\end{tabular}
&   24 & 16 & 16777216& 1054756
\\\hline
Triangular 4$\times$6 & 24 & 96 & 16777216 & 180070
\\\hline
Square 5$\times$5 & 25 & 200 & 33554432 & 172112
\\\hline
Buckyball  &	60   &  120  &
\begin{tabular}{c} 
1152921504606846976\\
$\approx %1.15\times
10^{18}$
\end{tabular}
& 
\begin{tabular}{c} 
9607679885269312\\
$\approx %0.96\times
10^{16}$
\end{tabular}
\\\hline
		\end{tabular}
\end{table}
\par
\section{Computer Program and Its Functionality}
We have written a C program to study different properties of deterministic  and statistical
lattice systems exploiting their symmetries.
Input of the program consists of the following elements:
\begin{itemize}
	\item
Graph $\lat=\left\{N_1,\ldots,N_n\right\}$.
$N_i$ is neighborhood of $i$th vertex,
i.e., the set of $k$ vertices adjacent to $i$th vertex.\\[-8pt]
\item \emph{Cellular automata branch:}\\
 Set of local rules $R=\left\{r_1,\ldots,r_m\right\}$. $r_i$ is integer number representing 
 bits of $i$th rule. The set $R$ includes the rules we are interested in. In particular,
 this set may contain only one rule (for detailed study).
\item \emph{Statistical models branch:}\\
Hamiltonian of the model.\\[-8pt]
 	\item
Some control parameters.
\end{itemize}
The program computes the automorphism group $\autlat$  and
\begin{itemize}
	\item in the case of cellular automata the program constructs \emph{phase portraits} 
	of automata \textit{modulo} $\autlat$ for
all rules from $R$.
\par
Manipulating the above mentioned control parameters we can
\begin{itemize}
	\item select automata with specified properties, for example, \emph{reversibility},
	\emph{conservation} of a given function on dynamical trajectories, etc.;
	\item search automata whose phase portraits contain specific structures, for example, 
 the	limit cycles of a given length,
	``\emph{gardens of Eden}'' \cite{GofL} or, more generally, isolated cycles, ``\emph{spaceships}'', etc.
\end{itemize}
\item in the case of statistical lattice model the program computes the partition function and other 
characteristics of the system, searches phase transitions.
\end{itemize}
\emph{Example of timing.}\\
The full run of all 136 symmetric 3-valent binary cellular automata on the dodecahedron (number of vertices = 20,
order of automorphism group = 120, number of states = 1048576, number of orbits = 9436) takes 
about 40 sec on a 1133MHz Pentium III personal computer.
\section{Deterministic Dynamical Systems}
In this section we point out a general principle of evolution of any causal dynamical system implied
by its symmetry, explain formation of soliton-like structures, and consider some results of computing with symmetric 3-valent cellular automata.
\subsubsection{Universal Property of Deterministic Evolution Induced by Symmetry.}
The splitting of the space $\stateset$ of functions on a lattice into the group orbits of different sizes imposes  \emph{universal restrictions} on behavior of a deterministic dynamical
system for any law that governs evolution of the system.
Namely, dynamical trajectories can obviously go only in the direction of \emph{non-decreasing sizes of orbits}.
In particular,  \emph{periodic trajectories} must lie \emph{within the orbits of the same size}.
Conceptually this restriction is an analog of the \emph{second law of thermodynamics} --- any isolated system may only lose information in its evolution.
\subsubsection{Formation of Soliton-like Structures.}
\label{solitonformation}
After some lapse of time the dynamics of finite discrete system is governed by its symmetry group,
that leads to appearance of \emph{soliton-like} structures. Let us clarify the matter.
Obviously phase portraits of the systems under consideration consist of attractors being limit cycles
and/or isolated cycles (including limit and isolated fixed points regarded as cycles of period one).
Now let us consider the behavior of the system which has come to a cycle, no matter whether the cycle is limit or isolated. The system runs periodically over some sequence of equal size orbits. 
The same orbit may occur in the cycle repeatedly. For example, the isolated cycle of period 6 in Fig. \ref{PhasePortrait} --- where a typical phase portrait \emph{modulo} automorphisms is presented --- passes through the sequence of orbits numbered%
\footnote{The program numbers orbits in the order of decreasing of their sizes and at equal sizes
the lexicographic order of lexicograhically minimal orbit representatives is used.}
as  0, 2, 4, 0, 2, 4, i.e., each orbit appears twice in the cycle.
\par
Suppose a state $\varphi(x)$ of the system running over a cycle belongs to $i$th orbit at some
moment $t_0$: $\varphi(x)\in O_i$.
At some other moment $t$ the system appears again in the same orbit with the state $\varphi_{t}(x) = A_{t_0t}\left(\varphi(x)\right)\in O_i$. Clearly, the evolution operator $A_{t_0t}$ can be replaced
by the action of some group element $g_{t_0t}\in\autlat$
\begin{equation}
\varphi_{t}(x) = A_{t_0t}\left(\varphi(x)\right)	= \varphi\left(g_{t_0t}^{-1}x\right).
\label{evogroup}
\end{equation}
The element $g_{t_0t}$ is determined uniquely \emph{modulo} subgroup 
$$\mathrm{Aut}\left(\lat; \varphi(x)\right)\subseteq\autlat$$ 
fixing the state $\varphi(x)$.
Equation (\ref{evogroup}) means that the initial cofiguration (shape) $\varphi(x)$ is completely reproduced
after some movement in the space $\lat$. 
Such soliton-like structures are typical for cellular automata. 
They are called ``\emph{spaceships}'' in the cellular automata community.
\par
\begin{figure}[!h]
\centering
\includegraphics[width=0.9\textwidth]{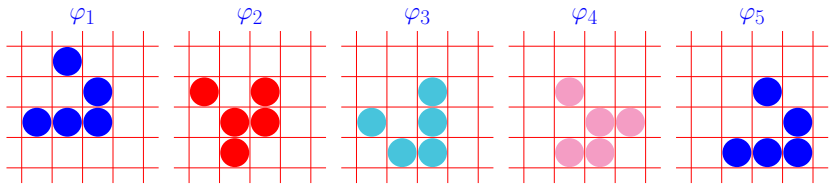}
\caption{Glider over translation group $\tra$ is cycle in \emph{four} group orbits.}
	\label{gliderT2}
\end{figure}
\begin{figure}[!h]
\centering
\includegraphics[width=0.9\textwidth]{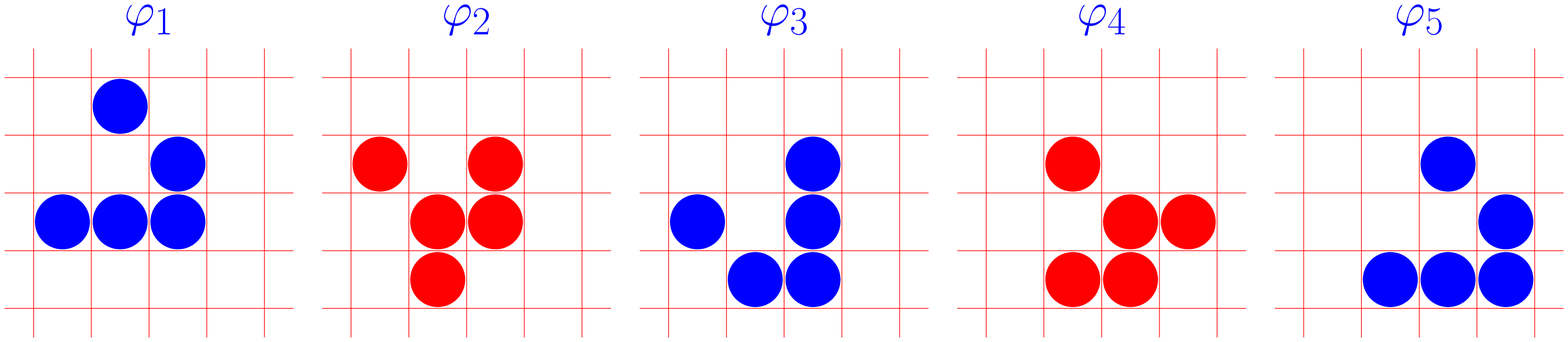}
\caption{Glider over maximal symmetry group $\tra\rtimes\dih$ is cycle in 
\emph{two} group orbits.}
	\label{gliderT2D4}
\end{figure}

Let us illustrate the group nature of such moving self-reproducing structures by the example
of ``\emph{glider}'' --- one of the simplest spaceships of Conway's automaton ``Life''.
This configuration moves along the diagonal of square lattice reproducing itself with one step
diagonal shift after four steps in time. 
If one considers only translations as a symmetry group of the lattice, then, as it is clear from 
Fig. \ref{gliderT2}, $\varphi_5$ is the first configuration lying in the same orbit%
\footnote{In Figs. \ref{gliderT2} and \ref{gliderT2D4} the configurations
belonging to the same orbit have identical colors.}
with $\varphi_1$,
i.e., for the translation group $\mathbf{T}^2$ glider is a cycle running over \emph{four} orbits.
\par
Our program constructs the maximum possible automorphism group for any lattice. 
For an $N\times N$ square toric lattice this group is the above mentioned $G_{N\times N}$ (we assume $N\neq4$,
see formula (\ref{normalgnn}) and subsequent discussion). 
\par
Now the glider is reproduced after two steps in time. 
As one can see 
from Fig. \ref{gliderT2D4}, $\varphi_3$ is obtained from $\varphi_1$ and $\varphi_4$ from $\varphi_2$ 
by combinations of translations, $90^o$ rotations and reflections. 
Thus, the glider in torus (and in the discrete plane obtained from the torus
as $n\to\infty$) is a cycle located in two orbits of maximal automorphism group.
\par
Note also that similar behavior is rather typical for continuous systems too.
Many equations of mathematical physics have solutions in the form of running wave
$\varphi\left(x-vt\right)$ 
$~\left(=  \varphi\left(g_t^{-1}x\right) ~\mbox{for Galilei group}\right)$.
One can see also an analogy between ``\emph{spaceships}'' of cellular automata and \emph{solitons} of 
KdV type equations.
The solitons --- like shape preserving moving structures in cellular automata --- are often
arise for rather arbitrary initial data.
\subsubsection{Cellular Automata with Symmetric Local Rules.}
As a specific class of discrete dynamical systems, we consider `one-time-step' cellular automata on
$k$-valent lattices with local rules symmetric with respect to all permutations of $k$
outer vertices of the neighborhood. This symmetry property is an immediate discrete analog
 of general local diffeomorphism invariance of fundamental physical theories based on continuous space.
The diffeomorphism group $\mathrm{Diff}(M)$ of the \emph{manifold} $M$ is very special subgroup of the infinite symmetric group $\mathrm{Sym}(M)$ of the \emph{set} $M$. 
\par	
%\noindent
As we demonstrated in \cite{KornyakCASC06}, in the binary case,
i.e., if the number of vertex values $q=2$, the automata with symmetric local rules are completely 
equivalent to generalized Conway's ``Game of Life'' automata \cite{GofL} and, hence, their 
rules can be formulated in terms of ``Birth''\!/``Survival'' lists.
\par
Adopting the convention that
the outer points and the root point of the neighborhood are denoted $x_1,\ldots,x_k$ and $x_{k+1}$,
respectively, we can write a \emph{local rule} determining one-time-step evolution of the root
in the form
\begin{equation}
x^{\prime}_{k+1} = f\left(x_1,\ldots,x_k,x_{k+1}\right).
	\label{localrule}
\end{equation}
The total number of rules (\ref{localrule}) symmetric with respect to permutations of points $x_1,\ldots,x_k$ is equal to
$
q^{\binom{k+q-1}{q-1}q}.
$
For the case of our interest ($k=3$, $q=2$) this number is 256.
\par 
It should be noted that the rules obtained from each other by
permutation of $q$ elements in the set $Q$ are equivalent since such permutation means nothing but renaming of
values. Thus, we can reduce the number of rules to consider. The reduced number can be counted
via \emph{Burnside's lemma} as a number of orbits of rules (\ref{localrule}) under the action of 
the group $\mathrm{S}_{q}$. The concrete expression depends on the cyclic structure of elements of
$\mathrm{S}_{q}$. For the case $q=2$ this gives the following number of non-equivalent rules
$$
	N_{rules}=2^{2k+1}+2^k.
$$
Thus, studying 3-valent binary case, we have to consider 136 different rules.
\subsubsection{Example of Phase Portrait.  Cellular Automaton 86.}
\begin{figure}[!h]
\centering
\includegraphics[width=\textwidth]{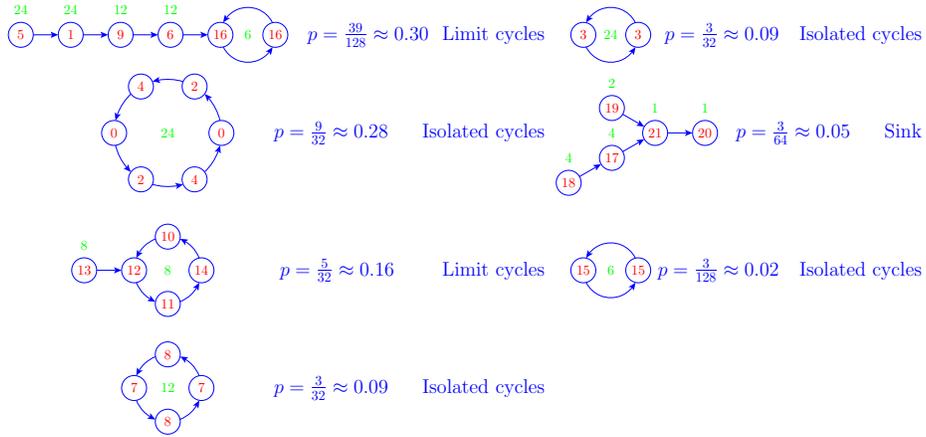}%
\caption{Rule 86. Equivalence classes of trajectories on hexahedron.
36 of 45 cycles are ``\emph{spaceships}''.}
	\label{PhasePortrait}
\end{figure}

As an example consider the rule 86 on hexahedron.
The number 86 is the ``little endian'' representation of the bit string 01101010 taken from
the last column of the rule table with 
$\mathrm{S}_3$-symmetric 
combinations of values for $x_1, x_2, x_3$
\begin{center}
\begin{tabular}[t]{cccc|l}
$x_1$&$x_2$&$x_3$&$x_4$&$x'_4$
\\
\hline
0&0&0&0&$0$\\
0&0&0&1&$1$\\
1&0&0&0&$1$\\
1&0&0&1&$0$\\
1&1&0&0&$1$\\
1&1&0&1&$0$\\
1&1&1&0&$1$\\
1&1&1&1&$0$\\
\end{tabular}
\\[-1pt]~~~~~~~~~~~~~~~~~~~~~.
\end{center}
The rule can also be represented in the ``Birth''\!/``Survival'' notation as B123/S0,
or as polynomial over the Galois field $\mathbb{F}_2$ (see \cite{KornyakCASC06})
$$
x'_4 = x_4+\sigma_3+\sigma_2+\sigma_1\enspace,
$$
where $\sigma_1 = x_1+x_2+x_3,\ \sigma_2 = x_1x_2+x_1x_3+x_2x_3,\ \sigma_3 = x_1x_2x_3$ are
\emph{symmetric} functions.
In Fig. \ref{PhasePortrait} the group orbits are represented by circles. The ordinal numbers of orbits are placed within these circles. The numbers over orbits and within cycles are sizes of the orbits
(recall that all orbits included in one cycle have the same size). 
The rational number $p$ indicates the \emph{weight} of the corresponding element of phase portrait. In other words, $p$ is
a probability to be in an isolated cycle or to be caught by an attractor at random choice of state: 
$p$ = (\emph{size of basin})/(\emph{total number of states}). Here \emph{size of basin} is sum of sizes of orbits involved in the struture.
\par
Note that most of cycles in Fig. \ref{PhasePortrait}  (36 of 45 or 80\%) are ``\emph{spaceships}''. Other computed examples also
confirm that soliton-like moving structures are typical for cellular automata.
\par
Of course, in the case of large lattices it is impractical to output full phase portraits
(the program easily computes tasks with up to hundreds thousands of different structures).
But it is not difficult to extract structures of interest, e.g., ``\emph{spaceships}'' or ``\emph{gardens of Eden}''.
\par
\subsubsection{Search for Reversibility.}
The program is able to select automata with properties specified at input.
One of such important properties is \emph{reversibility}. 
\par
\begin{figure}%[!h]
\centering
\includegraphics[width=0.7\textwidth]{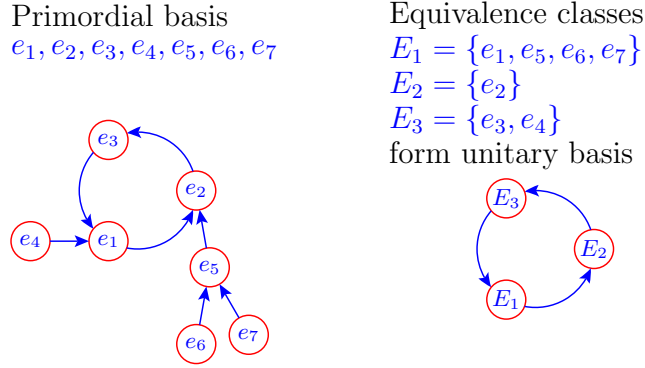}%
\caption{Transition from primordial to unitary basis.}
	\label{primtounit}
\end{figure}

In this connection we would like to mention recent works of G. 't Hooft.
One of the difficulties of Quantum Gravity is a conflict between irreversibility of Gravity
 --- information loss (dissipation) at the black hole horizon --- with reversibility and unitarity
of the standard Quantum Mechanics. In several papers 
of recent years (see, e.g., \cite{tHooft99,tHooft06}) 't Hooft developed the approach aiming to 
reconcile both theories. The approach is based on the following assumptions
 \begin{itemize}
	\item physical systems have \emph{discrete degrees of freedom} at tiny (Planck) distance scales;
	\item the states of these degrees of freedom form \emph{primordial} basis
	of Hilbert space (with nonunitary evolution);
	\item primordial states form \emph{equivalence classes}: two states are equivalent if they
	evolve into the same state after some lapse of time;
	\item the equivalence classes by construction form basis of Hilbert space with unitary
	 evolution described by time-reversible Schr\"odinger equation.
\end{itemize}
In our terminology this corresponds to transition to limit cycles: in a
finite time of evolution the limit cycle becomes physically indistinguishable from reversible isolated cycle
--- the system ``forgets'' its pre-cycle history. 
Fig. \ref{primtounit} illustrates construction of unitary Hilbert space from primordial.
\par
This irreversibility hardly can be found experimentally (assuming,
of course, that considered models can be applied to physical reality).
The system should probably spend time of order the Planck one 
($\approx10^{-44}$ sec) out of a cycle and potentially infinite time on the cycle. 
Nowadays, the shortest experimentally fixed time 
is about $10^{-18}$ sec or $10^{26}$ Planck units only.
\par
Applying our program to all 136 symmetric 3-valent automata we have the following.
There are two rules trivially reversible on all lattices 
\begin{itemize}
	\item  ~85 $\sim$ B0123/S $\sim$ $x'_4=x_4+1$,
	\item 170 $\sim$ B/S0123 $\sim$ $x'_4=x_4$.
\end{itemize}
Besides these uninteresting rules there are 6 reversible rules on \emph{tetrahedron}
\begin{itemize}
    \item ~43
     $~\sim~$ B0/S012  $~\sim~
     x'_4=x_4\left(\sigma_2+\sigma_1\right)+\sigma_3+\sigma_2+\sigma_1+1$,
    \item ~51
    $~\sim~$  B02/S02 $~\sim~ x'_4=\sigma_1+1$,
    \item ~77
    $~\sim~$ B013/S1  $~\sim~ x'_4=x_4\left(\sigma_2+\sigma_1+1\right)+\sigma_3+\sigma_2+1$,
    \item 178
    $~\sim~$ B2/S023  $~\sim~ x'_4=x_4\left(\sigma_2+\sigma_1+1\right)+\sigma_3+\sigma_2$,
    \item 204
    $~\sim~$ B13/S13  $~\sim~ x'_4=\sigma_1$,
    \item 212
    $~\sim~$ B123/S3 $~\sim~x'_4=x_4\left(\sigma_2+\sigma_1\right)+\sigma_3+\sigma_2+\sigma_1$.
\end{itemize}
Note that all these reversible rules are symmetric with respect to permutation of values 
$Q =\left\{0,1\right\}$. 
Two of the above rules, namely 51 and 204, are reversible on \emph{hexahedron} too.
There are no nontrivial reversible rules on all other lattices from Fig. \ref{lattices}.
Thus we may suppose that 't Hooft's picture is typical for discrete dynamical systems.
\section{Statistical Lattice Models and Mesoscopic Systems}
\label{mesosect}
\subsubsection{Statistical Mechanics.}
The state of deterministic dynamical system at any point of time is determined
uniquely by previous states of the system. A Markov chain --- for which 
transition from any state to any other state is possible with some probability --- is 
a typical example of \emph{non-deterministic} dynamical system. 
In this section we apply symmetry approach 
to the lattice models in statistical mechanics. These models can be regarded as special instances of Markov chains. 
\emph{Stationary distributions} of these Markov chains are studied by the methods of statistical mechanics. 
\par
The main tool of conventional statistical mechanics is the Gibbs \emph{canonical ensemble} --
imaginary collection of identical systems placed in a huge thermostat with temperature $T$. The statistical properties
of canonical ensemble are encoded in the \emph{canonical partition function}
\begin{equation}
Z=\sum\limits_{\sigma\in \stateset}\mathrm{e}^{-E_\sigma/k_B T}\enspace.
\label{cpf}
\end{equation} 
Here $\stateset$ is the set of microstates, $E_\sigma$ is energy of microstate $\sigma$, $k_B$ is Boltzmann's constant. 
The canonical ensemble is essentially asymptotic concept: its formulation is based on 
approximation called 
``thermodynamic limit''. For this reason, the canonical ensemble approach is applicable only to large 
(strictly speaking, infinite) homogeneous systems.
\subsubsection{Mesoscopy.}
Nowadays much attention is paid to study systems which are too large for a 
detailed microscopic description but too small 
for essential features of their behavior to be 
expressed in terms of classical thermodynamics.
This discipline, often called \emph{mesoscopy},
covers wide range of applications from nuclei, 
atomic clusters, nanotechnological structures to 
multi-star systems \cite{Imry02,Gross01,Gross04}.
To study \emph{mesoscopic} systems one should use 
more fundamental \emph{microcanonical ensemble} instead of canonical one.
A microcanonical ensemble is a collection of identical isolated systems at fixed energy.
Its definition does not include any approximating assumptions. In fact, the only key assumption 
of a microcanonical ensemble is that all its microstates are equally probable. This leads
to the \emph{entropy} formula
\begin{equation}
S_E=k_{B}\ln\staten{E}\enspace,
\label{entropy}
\end{equation}
or, equivalently, to the \emph{microcanonical partition function}
\begin{equation}
\staten{E}=\mathrm{e}^{S_E/k_{B}}\enspace.
\label{mcpf}
\end{equation}
Here $\staten{E}$ is the number of microstates at fixed energy $E$. 
In what follows we will omit Boltzmann's constant assuming $k_{B}=1$.
Note that in the thermodynamic limit the microcanonical and canonical
descriptions are equivalent and the link between them is provided by
the Laplace transform. On the other hand, mesoscopic systems demonstrate observable experimentally 
and in computation peculiarities of behavior like heat flows from cold to hot, negative specific heat 
or ``convex intruders'' in the entropy versus energy diagram,  etc. These anomalous -- from the point 
of view canonical thermostatistics -- features have natural explanation within  microcanonical statistical mechanics \cite{Gross04}.
\subsubsection{Lattice Models.}
In this section we apply symmetry analysis to study mesoscopic lattice
models. Our approach is based on exact enumeration of group orbits of microstates.
Since statistical studies are based essentially on different simplifying assumptions, 
it is important to control these assumptions by exact computation, wherever possible.
Moreover, we might hope to reveal with the help of exact computation subtle details
of behavior of system under consideration. 
\par
As an example, let us consider the Ising model. 
The model consists of \emph{spins} placed on a lattice. The set of vertex values is $Q=\set{-1,1}$ 
and the interaction Hamiltonian
is given by
\begin{equation}
H=-J\sum\limits_{(i,j)}s_is_j - B\sum\limits_{i}s_i\enspace,
\label{hamising}
\end{equation} 
where $s_i,s_j\in Q$; $J$ is a coupling constant ($J > 0$ and $J < 0$ correspond to \emph{ferromagnetic} 
and \emph{antiferromagnetic} cases, respectively);
the first sum runs over all edges $(i,j)$ of the lattice;
$B$ is an external ``magnetic'' field. The second sum $M = \sum\limits_{i}s_i$ is called the \emph{magnetization}. 
To avoid unnecessary technical details we will consider only the case $J > 0$
(assuming $J=1$) and $B=0$ in what follows.
\par
Since Hamiltonian and magnetization are constants on the group
orbits, we can count numbers of microstates corresponding to particular
values of these functions -- and hence compute all needed statistical characteristics -- 
simply by summation of sizes of appropriate orbits.
\par
\begin{figure}[!h]
\vspace*{-10pt}
\centering
\includegraphics[width=\textwidth]{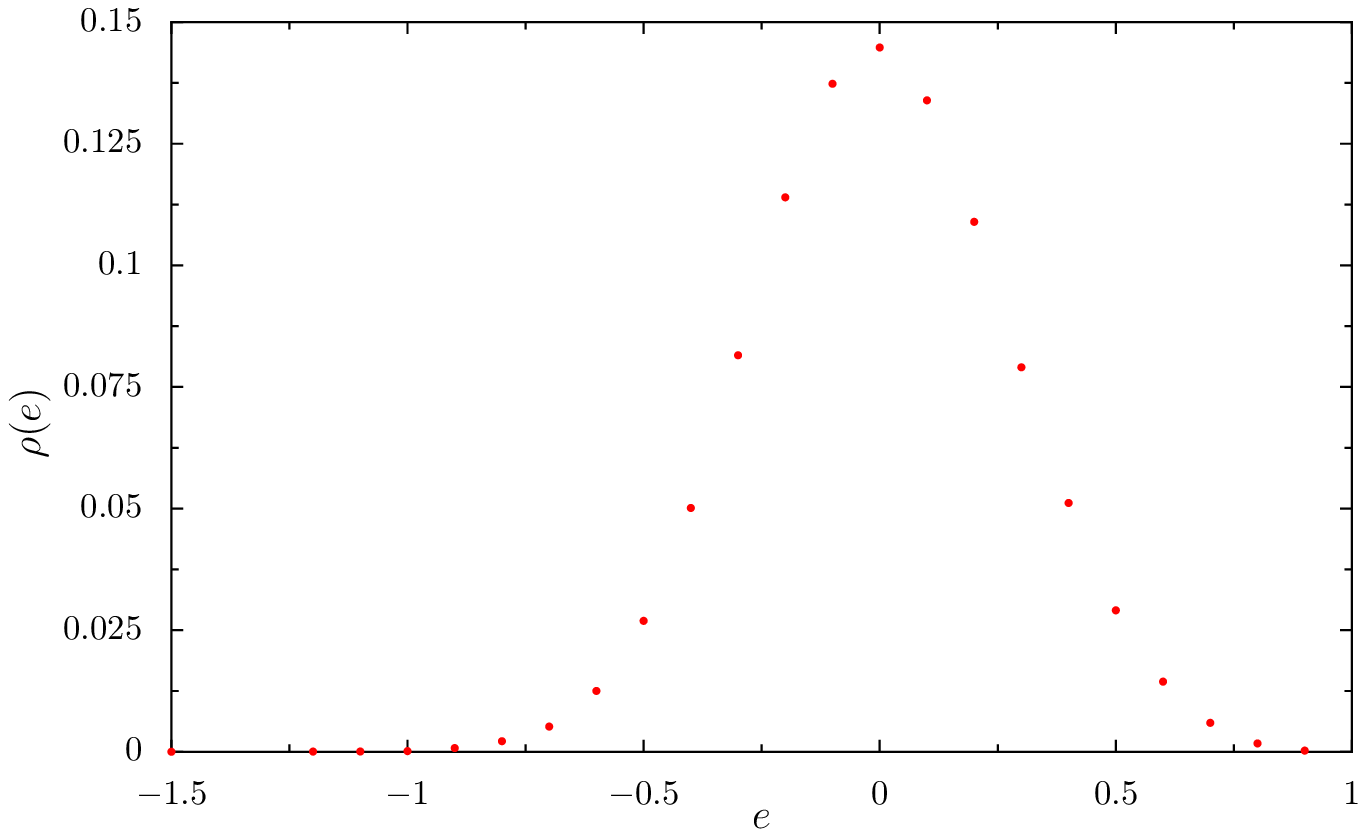}
\caption{Microcanonical density of states~ $\rho(e)=\staten{E}/\staten{}$ ~versus energy per vertex 
$e=E/\latvn$ for the Ising model on dodecahedron.} 
	\label{micro-part-dodeca}
\end{figure}

Fig. \ref{micro-part-dodeca} shows microcanonical partition function
for the Ising model on dodecahedron. Here total number of microstates
$\staten{}=1048576$, number of lattice vertices $\latvn=20$, energy $E$ is value of Hamiltonian.
\par
\begin{figure}[!h]
\vspace*{-10pt}
\centering
\includegraphics[width=\textwidth]{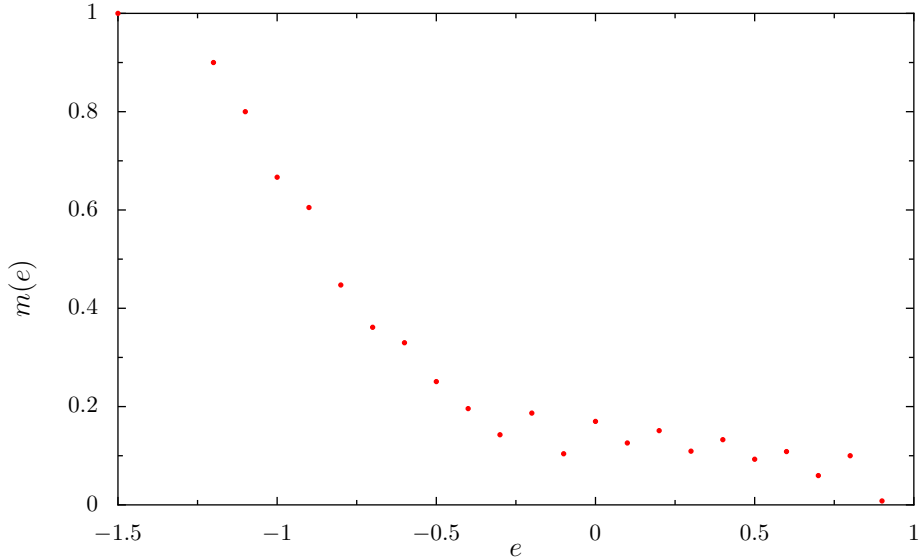}
\caption{Specific magnetization $m(e)=M(E)/\latvn$ vs. energy per vertex $e$ for the Ising model on dodecahedron.} 
	\label{magnetization-dodeca}
\end{figure}

%\noindent
Of course, other characteristics of the system can be computed easily via counting sizes of 
group orbits. For example, the magnetization is shown in Fig. \ref{magnetization-dodeca}.

\subsubsection{Phase Transitions.} Needs of nanotechnological science and nuclear physics attract special
attention to phase transitions in finite systems. Unfortunately
classical thermodynamics and the rigorous theory of critical phenomena in homogeneous infinite 
systems fails at the mesoscopic level. Several approaches have been proposed to identify phase
transitions in mesoscopic systems. Most accepted of them is search of \emph{``convex intruders''} \cite{GrossVotyakov} in
the entropy 
%$S$ 
versus energy 
%$E$ 
diagram. In the standard thermodynamics there is a relation
\begin{equation}
\left.\frac{\partial^2S}{\partial E^2}\right|_V = -\frac{1}{T^2}\frac{1}{C_V}\enspace,
\label{d2sde2}
\end{equation}
where $C_V$ is the specific heat at constant volume. It follows from (\ref{d2sde2}) that
$\left.\partial^2S/\partial E^2\right|_V < 0$ and hence the entropy versus energy diagram must be concave. 
Nevertheless, in mesoscopic systems there might be intervals of energy where $\left.\partial^2S/\partial E^2\right|_V > 0$.
 These intervals correspond to first-order 
phase transitions and are called \emph{``convex intruders''}. 
From the point of view of standard thermodynamics one can say about phenomenon 
of \emph{negative heat capacity}, of course, if one accepts that it makes sense 
to define the variables $T$ and $C_V$ as temperature and the specific heat at these
circumstances. In \cite{IspolatovCohen} it was demonstrated via computation with 
exactly solvable lattice models that the convex intruders flatten and disappear in 
the models with local interactions as the lattice size grows, while in the case of 
long-range interaction these peculiarities survive
even in the limit of an infinite system (both finite and long-range interacting 
infinite systems are typical cases of systems called \emph{nonextensive} in statistical mechanics).
\par
A convex intruder can be found easily by computer for the discrete systems we discuss here. 
Let us consider three adjacent values of energy $E_{i-1}, E_{i}, E_{i+1}$ and corresponding numbers 
of microstates
$\staten{E_{i-1}}, \staten{E_{i}}, \staten{E_{i+1}}$. In our discrete case the ratio 
$\left(E_{i+1}-E_{i}\right)/\left(E_{i}-E_{i-1}\right)$ is always rational number $p/q$ and
we can write the convexity condition for entropy in terms of numbers of microstates as easily computed
inequality
\begin{equation}
\staten{E_{i}}^{p+q} < \staten{E_{i-1}}^p\staten{E_{i+1}}^q\enspace.
\label{convcond}
\end{equation}
As a rule $E_{i+1}-E_{i}=E_{i}-E_{i-1}$ and inequality (\ref{convcond}) takes the form 
$$
\staten{E_{i}}^2 < \staten{E_{i-1}}\staten{E_{i+1}}\enspace.
$$
This form means that within convex intruder the number of states with the energy $E_{i}$ 
is less than \emph{geometric mean} of numbers of
states at the neighboring energy levels.
\par
\begin{figure}[!h]
\vspace*{-10pt}
\centering
\includegraphics[width=0.5\textwidth]{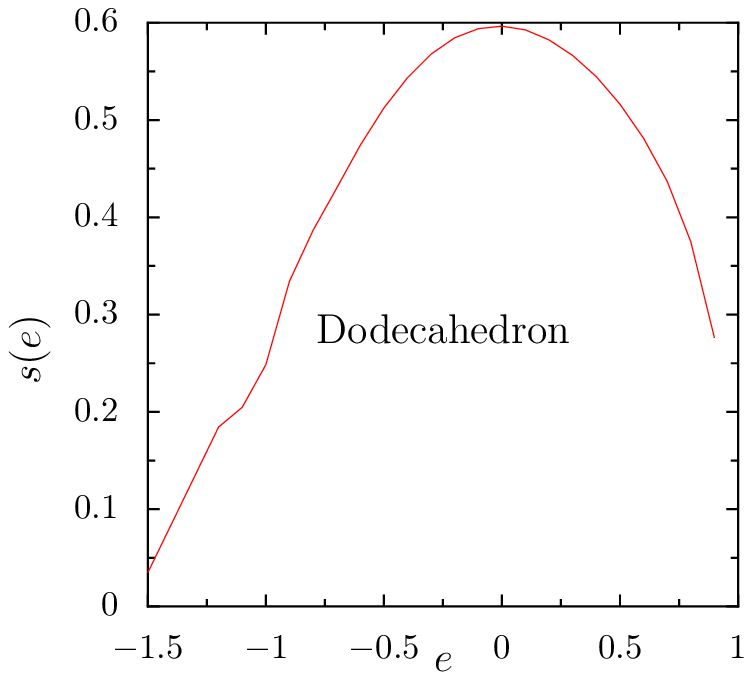}%
\includegraphics[width=0.5\textwidth]{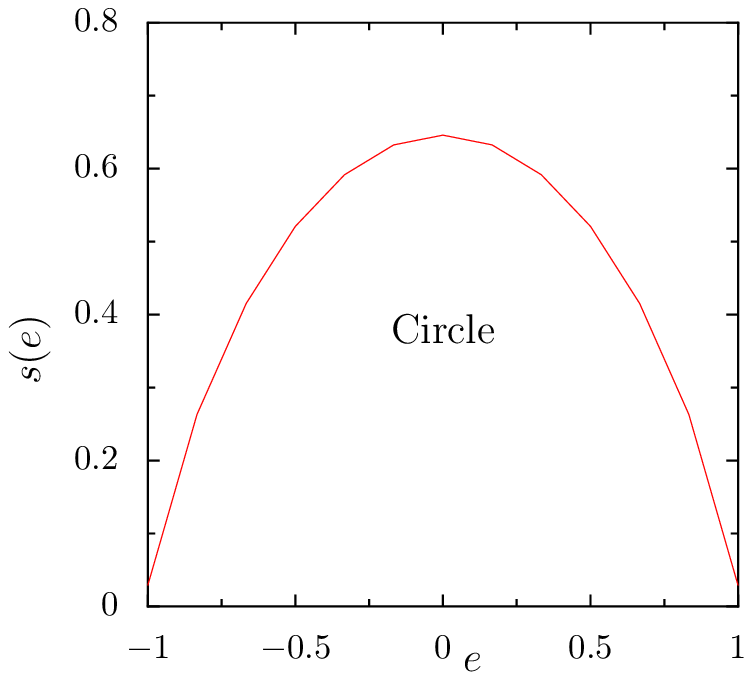}
\caption{Specific microcanonical entropy $s(e)=\ln\left(\staten{E}\right)/\latvn$ vs. 
energy per vertex $e$ for the Ising model on dodecahedron (\emph{left}) and on circle of length 24
 (\emph{right}). Left diagram contains distinct convex intruder in the interval $-1.2 \leq e \leq -0.9$ 
 and subtle one
in the interval $-0.8 \leq e \leq -0.6$. Right diagram is fully concave: one-dimensional Ising model has no
phase transitions.} 
	\label{micro-entropy-dodeca}
\end{figure}

Fig. \ref{micro-entropy-dodeca} shows the entropy vs. energy diagram for the Ising model on dodecahedron. 
The diagram has apparent convex intruder in the energy interval $\left[-24,-18\right]$.
Exact computation reveals also a subtle convex intruder in the interval $\left[-16,-12\right]$. 
(In terms of specific energy, as in Fig. \ref{micro-entropy-dodeca}, these intervals are
$\left[-1.2,-0.9\right]$ and $\left[-0.8,-0.6\right]$, respectively.) 
It is well known that one-dimensional
Ising model has no phase transitions. To illustrate the difference between the diagrams for 
the cases with and without phase transitions, we place also in Fig. \ref{micro-entropy-dodeca} the diagram
for Ising model on the 1D circle lattice with 24 vertices.
\par
\begin{figure}[!h]
\vspace*{-10pt}
\centering
\includegraphics[width=\textwidth]{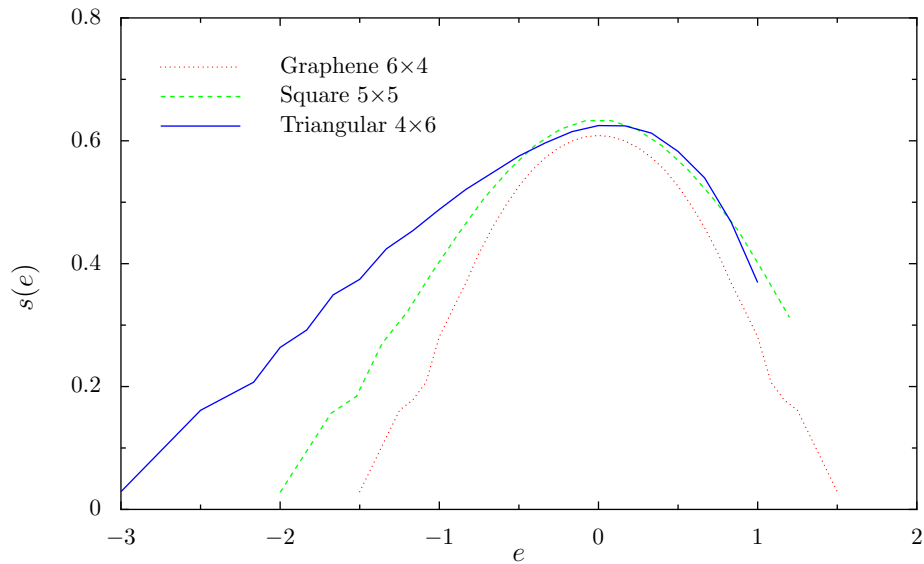}
\caption{Specific microcanonical entropy for the Ising model on 3-valent (\emph{dot} line, 
24 vertices), 4-valent (\emph{dash} line, 25 vertices) and 6-valent (\emph{solid} line, 24 vertices) tori.}
	\label{micro-entropy-tori}
\end{figure}

In Fig. \ref{micro-entropy-tori} we show the entropy-energy diagrams for lattices of different valences, 
namely, for 
3-, 4- and 6-valent tori. These lattices are marked in Fig. \ref{lattices} as ``Graphene 6$\times$4'', 
``Square 5$\times$5'' and ``Triangular 4$\times$6'', respectively.  The diagram for 3-valent torus 
is symmetric with respect to change sign of energy and contains
two pairs of adjacent convex intruders. 
One pair lies in the $e$-interval $[-1.25,-0.75]$ and another pair lies symmetrically in $[0.75,1.25]$. 
The 4-valent torus diagram contains two intersecting convex intruders in the intervals 
$[-1.68,-1.36]$ and $[-1.36,-1.04]$. The 6-valent torus diagram contains a whole cascade 
of 5 intersecting or adjacent intruders. Their common interval is $[-2.5,-0.5]$.
\section{Summary}
\begin{itemize}
	\item 
 A	C program for symmetry analysis of finite discrete dynamical systems has been created.
	\item We pointed out that trajectories of any deterministic dynamical system go always
	in the direction of nondecreasing sizes of group orbits. Cyclic trajectories run within orbits
	of the same size.
	\item
	After finite time evolution operators of dynamical system can be reduced to group actions.
	This lead to formation of moving soliton-like structures --- ``\emph{spaceships}'' in the case of
	cellular automata. Computer experiments show that ``\emph{spaceships}'' are typical for cellular automata.
	\item 
	Computational results for cellular automata with symmetric local rules
	allow to suppose that reversibility is 	rare property for discrete dynamical systems,
	and reversible systems are trivial.
	\item 
	We demonstrated capability of exact computing based on symmetries in search of phase transitions 
	for mesoscopic models in statistical mechanics.
\end{itemize}
\vspace*{-10pt}
\subsubsection*{Acknowledgments.}
I would like to thank Vladimir Gerdt whose comments improved
the presentation significantly.
This work was supported in part by the 
grants
07-01-00660 
from the Russian Foundation for Basic Research
 and
5362.2006.2
 from the Ministry of Education and Science of the Russian Federation.

%\fi
\end{document}